\documentclass[intlimits,twoside,a4paper]{article}
\usepackage{epsf}
\usepackage{amsmath,amssymb}
\usepackage{graphicx}
\usepackage[T2A]{fontenc}
\usepackage[cp1251]{inputenc}

\usepackage[eqsecnum]{cmpj2}

\newcommand{\mb}[1]{ { \mbox{\boldmath{$#1$}}}  }

\issue{2016}{19}{1}{13701}
\doinumber{10.5488/CMP.19.13701}

\title[Quasiparticle states due to preformed pairs]%
{Quasiparticle states driven by a scattering \\on the preformed electron pairs%
\thanks{This work is dedicated to professor Stefan Soko\l owski
on the occasion of his 65-th birthday.}}
\author{T. Doma\'nski}
\address{Institute of Physics, M. Curie-Sk\l odowska University,
              20-031 Lublin, Poland}

\date{Received November 13, 2015, in final form November 27, 2015}
\authorcopyright{T. Doma\'nski, 2016}

\begin{document}

\maketitle

\begin{abstract}
We analyze evolution of the single particle excitation spectrum of
the underdoped cuprate superconductors near the anti-nodal region,
considering temperatures below and and above the phase transition.
We inspect the phenomenological self-energy that reproduces the
angle-resolved-photoemission-spectroscopy (ARPES) data and we
show that above the critical temperature, such procedure implies
a transfer of the spectral weight from the Bogoliubov-type
quasiparticles towards the in-gap damped states. We also discuss
some possible microscopic arguments explaining this process.
\pacs 74.20.-z, 74.20.Mn, 74.40.+k
\end{abstract}

\section{Introduction}

Superconductivity (i.e., dissipationless motion of the charge carriers) is
observed at sufficiently low temperatures, when electrons from the vicinity
of the Fermi surface are bound in the pairs and respond collectively (rather
than individually) to any external perturbation such as electromagnetic field,
pressure, temperature gradient, etc. Depending on specific materials,
the pairing mechanism can be driven by phonons (in classical superconductors),
magnons (in heavy fermion compounds) or by the antiferromagnetic exchange
interactions originating from the Coulomb repulsion (in cuprate oxides).
In most cases, the electron pairs are formed at the critical value $T_\text{c}$,
marking the onset of superconductivity. There are, however, numerous
exceptions to this rule. For instance, in the cuprate superconductors
\cite{Kaminski2014} or in the ultracold fermionic gasses \cite{Jin-10},
such pairs pre-exist  well above $T_\text{c}$. To some extent, their presence
causes the properties reminiscent of the superconducting state.

Early evidence for the preformed pairs existing above $T_\text{c}$ has been
indicated in the muon scattering experiments \cite{Uemura}. Later on,
their existence was supported by the ultrafast (tera-Hertz) optical
spectroscopy \cite{Orenstein-99a,Orenstein-99b} and the large Nernst effect \cite{Ong-00a,Ong-00b}.
Spectroscopic signatures of the preformed pairs have been also detected
directly in the ARPES measurements on yttrium \cite{Argonne-08} and
lanthanum \cite{Villigen-08} cuprate oxides, revealing the Bogoliubov-type
quasiparticle dispersion above $T_\text{c}$. Furthermore, the STM imaging provided
clear fingerprints of such dispersive Bogoliubov quasiparticles (by the unique
octet patterns) being unchanged from temperatures deep in the superconducting
region up to $1.5 T_\text{c}$ \cite{Davis-09}. Superconducting fluctuations
above $T_\text{c}$ have been also reported by the Josephson-like tunneling
\cite{Bergeal-08} and the proximity effect induced in the nanosize metallic
slabs deposited on La$_{2-x}$Sr$_{x}$CuO$_{4}$ \cite{Yuli-09}. More recently,
the residual Meissner effect has been experimentally observed above the transition
temperature $T_\text{c}$ by the torque magnetometry \cite{Ong-10} and other
measurements \cite{Iye-10,Bernardi-10}. Additional evidence for the
superconducting-like behaviour above $T_\text{c}$ has been seen in the
high-resolution ARPES measurements \cite{Kondo2009}, the superfulid
fraction observed in the $c$-axis optical measurements
Re$\{ \sigma_\text{c}(\omega) \}$ \cite{Dubroka2011},
the Josephson spectroscopy for YBaCuO-LaSrCuO-YBaCuO junction
using LaSrCuO in the pseudogap state well above $T_\text{c}$ \cite{Kirzhner2014},
optical conductance in the pseudogap state of YBaCuO superconductor
\cite{Moon2014} and the photo-enhanced antinodal conductivity in
pseudogap state of the high $T_\text{c}$ superconductors \cite{Cilento2014}.

Preformed pairs are correlated above $T_\text{c}$ only on some short temporal
$\tau_{\phi}$ and spatial $l_{\phi}$  scales \cite{Franz-07,Senthil-09a,Senthil-09b}.
For this reason, the superconducting fluctuations are manifested in very
peculiar way \cite{deLlano2014}. Their influence on the single particle
spectrum is manifested by: a) two Bogoliubov-type branches and b) additional
in-gap states that are over-damped (have a short life-time). Temperature
has a strong effect on the transfer of the spectral weight between
these entities. In the underdoped cuprate superconductors, such
a transfer is responsible for filling-in the energy gap \cite{Kondo2009,STM},
instead of closing it (as in the classical superconductors).
Some early results concerning superconducting fluctuations
have been known for a long time \cite{Abrahams-70,Schmid-70}, but
they attracted much more interest in the context of cuprate superconductors
\cite{Abrahams-70,Schmid-70,Ranninger-95,Emery-95,Tchernyshyov-97,Norman-98a,Norman-98b,
Fujimoto-02,Domanski2003a,Domanski2003b,Chubukov-07} and ultracold fermion superfluids
\cite{Levin_K,Micnas2014}.

In this work, we study qualitative changeover of the single particle electronic
spectrum of the underdoped cuprate oxides for temperatures varying from below
$T_\text{c}$ (in the superconducting state) to above $T_\text{c}$ , where the preformed
pairs are not long-range coherent. In the superconducting state,
the usual Bogoliubov-type quasiparticles are driven by the Bose-Einstein
condensate of the (zero-momentum) \linebreak Cooper pairs. We find that above $T_\text{c}$,
the Bogoliubov quasiparticles are still preserved, but the scattering
processes driven by the finite momentum pairs  contribute the in-gap
states whose life-time substantially increases with increasing temperature.
We discuss this process on the phenomenological as well as microscopic
arguments. Roughly speaking, a feedback of the electron pairs on the unpaired electrons
resembles the long-range translational and orientational order that
develops between the amphiphilic particles in presence of
the ions at solid state surfaces studied by S. Soko\l owski with
coworkers \cite{Sokolowski2014}.

\section{Microscopic formulation of the problem}

To account for the coherent/incoherent pairing we consider the Hamiltonian
\begin{eqnarray}
\hat{H} =\sum_{{\bf k},\sigma} \left( \varepsilon_{\bf k}  -
\mu \right) \hat{c}_{{\bf k}\sigma}^{\dagger} \hat{c}_{{\bf k}
\sigma} + \frac{1}{N} \; \sum_{{\bf k},{\bf k}',{\bf q}}
V_{{\bf k},{\bf k}'}({\bf q})\hat{c}_{{\bf k}'\uparrow}^{\dagger}
\hat{c}_{{\bf q}-{\bf k}'\downarrow}^{\dagger}
\hat{c}_{{\bf q}-{\bf k}\downarrow}\hat{c}_{{\bf k}\uparrow}
\label{model}
\end{eqnarray}
describing the mobile electrons of kinetic energy $\varepsilon_{\bf k}$
(where $\mu$ is the chemical potential) interacting via the two-body
potential $V_{{\bf k},{\bf k}'}({\bf q})$. We assume a separable form
$V_{{\bf k},{\bf k}'} = - g \;\eta_{\bf k}\; \eta_{{\bf k}'}$ of this
pairing potential (with $g>0$).
In the nearly two-dimensional cuprate superconductors with the prefactor
\linebreak $\eta_{\bf k}=\frac{1}{2} \left[\cos{(a k_x)}+\cos{(a k_y)}\right]$
(where $a$ is the unit length in CuO$_{2}$ planar structure), such
pairing potential induces the $d$-wave symmetry order parameter
\cite{Matsui-03,Chatterjee-09}.

The  Hamiltonian (\ref{model}) can be recast in a more
compact form, by introducing the pair operators
\begin{eqnarray}
\hat{b}_{\bf q} = \frac{1}{\sqrt{N}} \sum_{\bf k} \eta_{\bf k}
\hat{c}_{{\bf q}-{\bf k}\downarrow} \hat{c}_{{\bf k}\uparrow}
\label{pair_operator}
\end{eqnarray}
and $\hat{b}_{\bf q}^{\dagger}=(\hat{b}_{\bf q})^{\dagger}$, when
the two-body interactions simplify to
\begin{eqnarray}
 \frac{1}{N} \; \sum_{{\bf k},{\bf k}',{\bf q}}
V_{{\bf k},{\bf k}'}({\bf q})\hat{c}_{{\bf k}\uparrow}^{\dagger}
\hat{c}_{{\bf q}-{\bf k}\downarrow}^{\dagger}
\hat{c}_{{\bf q}-{\bf k}'\downarrow}\hat{c}_{{\bf k}'\uparrow}
= - \sum_{\bf q} g \; \hat{b}_{\bf q}^{\dagger} \hat{b}_{\bf q} \,.
\label{model_bis}
\end{eqnarray}
Using the Heisenberg equation of motion ($\hbar = 1$)
\begin{eqnarray}
\ri\frac{\rd}{\rd t} \hat{c}_{{\bf k}\uparrow}=\left(\varepsilon_{\bf k}
 - \mu\right)\hat{c}_{{\bf k}\uparrow} - g\;\eta_{\bf k}
\frac{1}{\sqrt{N}} \; \sum_{\bf q} \hat{c}_{{\bf q}-{\bf k}
\downarrow}^{\dagger} \hat{b}_{\bf q}
\label{Heisenberg_eqn}
\end{eqnarray}
we immediately notice that the single-particle properties of this model (\ref{model}):
\begin{itemize}
\item[{a)}]  are characterized by the mixed particle and hole degrees
of freedom (because the annihilation operators $\hat{c}_{{\bf k}\uparrow}$
couple to the creation operators $\hat{c}_{{\bf q}-{\bf k} \downarrow}^{\dagger}$),
\item[{b)}] depend on the pairing field $\hat{b}_{\bf q}$ (appearing in
the equation of motion $\frac{\rd}{\rd t} \hat{c}_{{\bf k}\uparrow}$).
\end{itemize}
Both these features manifest themselves in the superconducting state,
when there exists the Bose-Einstein (BE) condensate $\langle \hat{b}_{{\bf q}
 = {\bf 0}} \rangle \neq 0$ of the  Cooper pairs. They
also survive in the normal state, as long as the preformed
(finite-momentum) pairs are present below the same characteristic temperature
$T^{*}$ marking an onset of the electron pairing. In the next section
we explore their role in the superconducting state ($T\leqslant T_\text{c}$) and
in the pseudogap region ($T_\text{c}<T<T^{*}$).

\section{Pairs as the scattering centers}

The Heisenberg equation of motion (\ref{Heisenberg_eqn}) indicates that
the electronic states are affected by the pairing field $\hat{b}_{\bf q}$.
Let us consider the generic consequences of such Andreev-type scattering,
separately considering: the BE condensed ${\bf q}  =  {\bf 0}$
and the finite-momentum ${\bf q} \neq {\bf 0}$ pairs.

\subsection{The effect of the Bose-Einstein condensed pairs}

We start by considering the usual BCS approach, when only the zero
momentum pairs are taken into account. This situation has a particularly
clear interpretation within the path integral formalism, treating the
pairing field via the Hubbard-Stratonovich transformation and determining
it from the minimization of action (the saddle point solution). The same
result can be obtained using the equation of motion (\ref{Heisenberg_eqn}),
focusing on the effect of ${\bf q}={\bf 0}$ pairs
\begin{eqnarray}
\ri\frac{\rd}{\rd t} \hat{c}_{{\bf k}\uparrow} & \simeq & \left(
\varepsilon_{\bf k}  - \mu\right)\hat{c}_{{\bf k}\uparrow}
- g\;\eta_{\bf k} \; \hat{c}_{-{\bf k}\downarrow}^{\dagger}
\frac{\hat{b}_{\bf 0}}{\sqrt{N}} ,
\label{BEC_term}
\\
\ri\frac{\rd}{\rd t} \hat{c}_{{\bf k}\downarrow}^{\dagger} & \simeq &
- \left(\varepsilon_{\bf k}  - \mu\right)\hat{c}_{{\bf k}
\downarrow}^{\dagger} - g\;\eta_{\bf k}  \frac{\hat{b}_{\bf 0}
^{\dagger}}{\sqrt{N}} \; \hat{c}_{{\bf k}\uparrow}.
\label{BEC_term_dagger}
\end{eqnarray}
Macroscopic occupancy of the ${\bf q} = {\bf 0}$ state implies that
the bosonic operators $\hat{b}_{\bf 0}^{(\dagger)}$ can be treated as
complex numbers $b_{\bf 0}^{(*)}$. By introducing
the order parameter
\begin{eqnarray}
\Delta_{\bf k}=g\eta_{\bf k} \frac{\langle \hat{b}_{\bf 0}\rangle}{\sqrt{N}}=g
\eta_{\bf k} \frac{1}{N} \sum_{{\bf k}'} \eta_{{\bf k}'} \langle
\hat{c}_{-{\bf k}'\downarrow} \hat{c}_{{\bf k}'\uparrow} \rangle
\label{gap}
\end{eqnarray}
the equations (\ref{BEC_term}), (\ref{BEC_term_dagger}) can be solved
exactly using the standard Bogoliubov-Valatin transformation. In such
BCS approach, the classical superconductivity has close analogy with the
superfluidity of weakly interacting bosons, whose collective sound-like
mode originates from the interaction between the finite-momentum bosons
and the BE condensate.

In the present context, the zero-momentum Copper pairs substantially
affect the single particle excitation spectrum (and the two-body
correlations as well). The single-particle Green's function ${\mb G}
({\bf k},\tau) =  - \hat{T}_{\tau} \langle \hat{c}_{{\bf k}\uparrow}(\tau)
\hat{c}_{{\bf k}\uparrow}^{\dagger} \rangle$, where $\hat{T}_{\tau}$
is the time ordering operator, obeys the Dyson equation
\begin{eqnarray}
\left[ {\mb G}({\bf k},\omega) \right]^{-1} =  \omega-
\varepsilon_{\bf k} + \mu - \Sigma({\bf k},\omega),
\label{Dyson}
\end{eqnarray}
with the BCS self-energy
\begin{eqnarray}
\Sigma({\bf k},\omega) =
\frac{|\Delta_{\bf k}|^{2}}{\omega+( \varepsilon_{\bf k}-\mu )} .
\label{BCS_selfenergy}
\end{eqnarray}
The self-energy (\ref{BCS_selfenergy}), accounting for the Andreev-type
scattering of the ${\bf k}$-momentum electrons on the Cooper pairs, can
be alternatively obtained from the bubble diagram.
The related spectral function \linebreak $A({\bf k},\omega)=-\pi^{-1} \text{Im}\,
{\mb G}({\bf k},\omega+\ri 0^{+})$ is thus characterized by the two-pole
structure
\begin{eqnarray}
A({\bf k},\omega) = u^{2}_{\bf k} \; \delta(\omega -
E_{\bf k}) + v^{2}_{\bf k} \; \delta(\omega + E_{\bf k})
\label{BCS_spectral}
\end{eqnarray}
with the Bogoliubov-type quasiparticle energies $E_{\bf k} =
\pm \sqrt{\left( \varepsilon_{\bf k}  - \mu \right)^{2} +
\Delta_{\bf k}^{2}}$ and the spectral weights $u_{\bf k}^{2}=
\frac{1}{2}\left[1 + (\varepsilon_{\bf k} - \mu)/E_{\bf k}
\right]$ and  $v_{\bf k}^{2} = 1-u_{\bf k}^{2}$. Let us remark
that these quasiparticle branches are separated by the (true)
energy gap $|\Delta_{\bf k}|$. In classical superconductors,
$\Delta_{\bf k}$ implies the off-diagonal-long-range-order (ODLRO)
that quantitatively depends on concentration of the BE condensed
Cooper pairs. ODLRO is responsible for a dissipationless motion
of the charge carriers and simultaneously causes the Meissner
effect via the spontaneous gauge symmetry breaking.

\subsection{The effect of the non-condensed pairs}

In this section we shall study effect of the finite-momentum
pairs existing above $T_\text{c}$, which no longer develop any ODLRO
because there is no BE condensate. Nevertheless, according to
(\ref{Heisenberg_eqn}), the single and paired fermions are still
mutually dependent. This fact suggests that the previous BCS form
(\ref{BCS_selfenergy}) should be replaced by some corrections
originating from the finite momentum pairs. Let us denote
the pair propagator by $L({\bf q},\tau) = - \; \hat{T}_{\tau}
\langle \hat{b}({\bf r},\tau) \hat{b}^{\dagger}({\bf 0},0) \rangle$
and assume its Fourier transform in the following form
\begin{eqnarray}
L({\bf q},\omega) = \frac{1}{
\omega - E_{\bf q} - \ri \; \Gamma({\bf q},\omega)},
\label{pair}
\end{eqnarray}
where $E_{\bf q}$ stands for the effective dispersion
of pairs and $\Gamma({\bf q},\omega)$ describes  the
inverse life-time. Taking into account the equation
(\ref{Heisenberg_eqn}), we express the self-energy via
the bubble diagram
\begin{eqnarray}
\Sigma({\bf k},\omega) = -T \sum_{\ri\nu_{n},{\bf q}}
\frac{1}{ \omega -\xi_{{\bf q}-{\bf k}} - \ri\nu_{n}}
\; L({\bf q},\ri\nu_{n}),
\label{L_pert}
\end{eqnarray}
where $\xi_{{\bf q}-{\bf k}}=\varepsilon_{{\bf q}-{\bf k}}-\mu$
and $\ri\nu_{n}$ is the bosonic Matsubara frequency. Since above
$T_\text{c}$ the preformed pairs are only short-range correlated
\cite{Franz-07,Senthil-09a,Senthil-09b}, we  impose
\begin{eqnarray}
\langle
\hat{L}^{\dagger}({\bf r},t)
\hat{L}({\bf 0},0)
\rangle \propto
\mbox{\rm exp}\left( - \frac{|t|}{\tau_{\phi}}
- \frac{|{\bf r}|}{\xi_{\phi}} \right)  .
\label{decay}
\end{eqnarray}
Following  T.~Senthil and P.A.~Lee \cite{Senthil-09a,Senthil-09b}, one can estimate
the single particle Green's function ${\cal{G}}({\bf k},\omega)$ using
the following interpolation
\begin{eqnarray}
\Sigma({\bf k},\omega)=  \Delta^{2} \frac{ \omega - \xi_{\bf k}}
{ \omega^{2}- \left( \xi_{\bf k}^{2}+\pi \Gamma^{2} \right) } ,
\label{interpol}
\end{eqnarray}
where $\Delta$ is the energy gap due to pairing and the other parameter
$\Gamma$ is related to damping of the subgap states. In the low energy
limit (i.e., for $|\omega| \ll \Delta$) the dominant contribution
comes from the in-gap quasiparticle whose residue is $Z\equiv
\left[ 1+ {\Delta^{2}}/({\pi \Gamma^2}) \right]^{-1}$, whereas
at higher energies the BCS-type quasiparticles are recovered.
This selfenergy (\ref{interpol}) can be derived from the microscopic
considerations \cite{Domanski2011} within the two-component model,
describing itinerant fermions coupled to the hard-core bosons
\cite{Friedberg-89,Micnas-90a,Micnas-90b,Geshkenbein-97,Auerbach-09a,Auerbach-09b,Rice-09,Ranninger-10}.

The other (closely relative) phenomenological ansatz \cite{Norman-98a,Norman-98b}
\begin{eqnarray}
\Sigma({\bf k},\omega) = \frac{\Delta^{2}}
{\omega + \xi_{\bf k} + \ri \Gamma_{0}} - \ri\Gamma_{1}
\label{selfenergy}
\end{eqnarray}
has been inferred considering the ``small fluctuations''  regime \cite{Abrahams-70}.
Experimental lineshapes of the angle resolved photoemission spectroscopy
obtained for the cuprate superconductors at various doping levels and
temperatures (including the pseudogap regime) amazingly well coincide
with this simple formula (\ref{selfenergy}).
The gap and the phenomenological parameters
$\Gamma_{0}$, $\Gamma_{1}$ are in general momentum-dependent,
but for a given direction in the Brillouin zone one can restrict
only to their temperature and doping variations. From now onwards
we shall focus on such antinodal region.

\begin{figure}[!t]
\epsfxsize=8cm\centerline{\epsffile{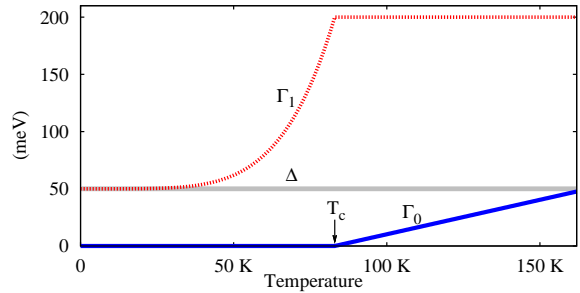}}
\caption{(Color online) Temperature dependence of the phenomenological
parameters $\Delta$, $\Gamma_0$ and $\Gamma_1$ which, through
the self-energy (\ref{selfenergy}), reproduce the experimental
profiles of the underdoped Bi2212 ($T_\text{c} = 83$~K) sample.
This fitting is adopted from references \cite{Norman-98a,Norman-98b}.}
\label{fig1}
\end{figure}

In the overdoped samples, $\Gamma_{0}$ can be practically discarded
from (\ref{selfenergy}) and the remaining parameter $\Gamma_{1}$
simply accounts for $T$-dependent broadening of the Bogoliubov peaks
until they disappear just above $T_\text{c}$. Physical origin of $\Gamma_{1}$
is hence related to the particle-particle scattering. On the contrary, in the
underdoped regime, there exists a pseudogap up to temperatures $T^{*}$
which by far exceed $T_\text{c}$. To reproduce the experimental lineshapes,
one must then incorporate the other parameter $\Gamma_{0}$ (nonvanishing
only above $T_\text{c}$) which is scaled by $T - T_\text{c}$ as shown in figure~\ref{fig1} reproduced from references \cite{Norman-98a,Norman-98b}.  Since $\Gamma_{0}$
enters the self-energy (\ref{selfenergy}) through the BCS-type structure,
its origin is related to the particle-hole scatterings.

We now inspect some consequences of the parametrization (\ref{selfenergy})
applicable for the pseudogap regime $T > T_\text{c}$ in the underdoped cuprates.
Since neither the magnitude of $\Gamma_{1}$ nor $\Delta$ seem to vary
over a large temperature region above $T_\text{c}$, it is obvious that
the qualitative changes are there dominated by scatterings in the
particle-hole channel, i.e., due to $\Gamma_{0}$. Roughly speaking,
these processes  are responsible for filling-in the low energy states
upon increasing $T$ as has been evidenced by ARPES \cite{Kondo2009}
and STM \cite{STM} measurements. On a microscopic level, such changes
can be assigned to scattering on the preformed pairs.

For analytical considerations, let us rewrite the complex
self-energy (\ref{selfenergy}) as
\begin{eqnarray}
\Sigma({\bf k},\omega) = \left( \omega + \xi_{\bf k} \right)
\frac{ \Delta^{2} }{\left( \omega + \xi_{\bf k}\right)^{2}
+ \Gamma_{0}^{2}} - \ri \Gamma_{\bf k}(\omega),
\end{eqnarray}
where the imaginary part is
\begin{eqnarray}
\Gamma_{\bf k}(\omega) = \Gamma_{\bf 1} + \Gamma_{0}
\frac{  \Delta^{2} }{\left( \omega + \xi_{\bf k}\right)^{2}
+ \Gamma_{0}^{2}} .
\end{eqnarray}
In what follows we indicate
that above $T_\text{c}$ the excitation spectrum can consist
of altogether three different states, two of them
corresponding to the Bogoliubov modes (signifying
particle-hole mixing characteristic for the superconducting state)
and another one corresponding to the single particle fermion
states which form inside the pseudogap.
These states start to appear at $T = T_\text{c}^{+}$ and
initially represent heavily overdamped modes containing
infinitesimal spectral weight (see reference \cite{Senthil-09a,Senthil-09b}
for a more detailed discussion). Upon increasing temperature,
their life-time gradually increases and simultaneously the
in-gap states absorb more and more spectral weight
at the expense of the Bogoliubov quasiparticles. Finally
(in the particular case considered here, this happens
roughly near $2T_\text{c}$) the single particle fermion
states become dominant.

\begin{figure}[!t]
\epsfxsize=8cm\centerline{\epsffile{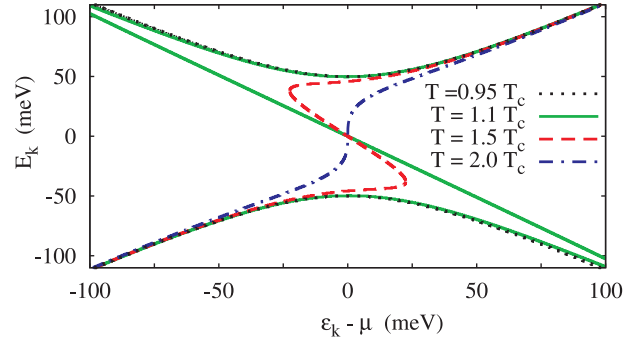}}
\caption{(Color online)
Dispersion of $\omega=E_{\bf k}$ representing the solutions
of equation (\ref{definition}) for  $T/T_\text{c}=0.95$ (dotted
line), $1.1$, $1.5$ and $2$ (solid curves as described) obtained
for the parameters used in references \cite{Norman-98a,Norman-98b}. We notice
three different crossings (\ref{constraint_for_poles}), two of
them related to the Bogoliubov modes and additional one
appearing in between.}
\label{poles_num}
\end{figure}

Anticipating the relevance of (\ref{selfenergy}) to the strongly
correlated cuprate materials, one can determine the single
particle Green's function ${\mb G}({\bf k},\omega)$ and
the corresponding spectral function $A({\bf k},\omega)$.
Quasiparticle energies are determined
by poles of the Green's function, i.e.,
\begin{eqnarray}
 \omega - \xi_{\bf k} - \text{Re} \left\{
 \Sigma({\bf k},\omega) \right\} =0
\label{definition}
\end{eqnarray}
provided that the imaginary part $\Gamma_{\bf k}(\omega)$
disappears. We clearly see that the latter requirement cannot
be satisfied for $\Gamma_1 \neq 0$ regardless of $\Gamma_0$.
Formally this means that the life-time of herein discussed
quasiparticles is not infinite. Let us check
these eventual (finite life-time) quasiparticle states determined
through (\ref{definition}). Using the self-energy (\ref{selfenergy}),
the condition (\ref{definition}) is equivalent to
\begin{eqnarray}
\left( \omega   -   \xi_{\bf k}\right) -  \left( \omega   +
  \xi_{\bf k} \right) \frac{ \Delta^{2} }{\left( \omega +
\xi_{\bf k}\right)^{2} + \Gamma_{0}^{2}} = 0 .
\label{constraint_for_poles}
\end{eqnarray}
In general, there  appear three solutions (figure~\ref{poles_num})
depending on temperature via the parameter $\Gamma_{0}$.

{\em Superconducting region}.
The fitting procedure  \cite{Norman-98a,Norman-98b} has estimated that
the parameter $\Gamma_{0} $ vanishes in the superconducting
state  $T  \leqslant  T_\text{c}$. Under such conditions,
(\ref{constraint_for_poles}) yields the standard BCS poles
at $E_{\bf k} = \pm \sqrt{\xi_{\bf k}^2 +\Delta^2}$. Due
to $\Gamma_{1} \neq 0$, they show up in the spectral
function $A({\bf k},\omega)$ as Lorentzians whose
broadening corresponds to the inverse life-time
of the Bogoliubov modes. Owing to $T$-dependence
of $\Gamma_{1}$  (see figure~\ref{fig1}), the
broadening of these peaks increases upon approaching
$T_\text{c}$ from below, albeit $A({\bf k},\omega = 0) = 0$.
Experimentally this process can be observed as the smearing
of the coherence peaks \cite{Kaminski2014}.

\begin{figure}[!b]
\epsfxsize=8cm\centerline{\epsffile{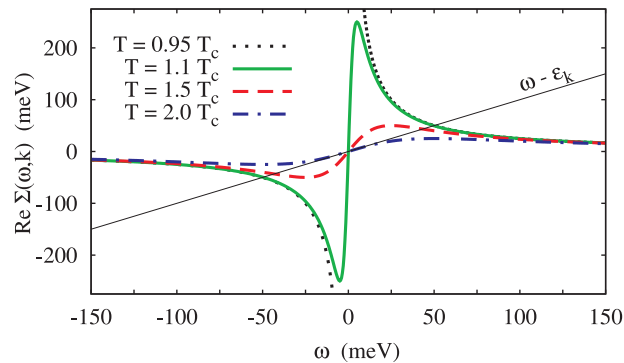}}
\caption{(Color online)
The real part of the self-energy
$\Sigma({\bf k},\omega)$ for $\varepsilon_{\bf k} = \mu$
at several representative temperatures $T/T_\text{c}$=$0.95$
(dotted line) and $1.1$, $1.5$, $2.0$ (as denoted). Below
$T_\text{c}$ there exist two poles at $\omega=\pm\Delta$
whereas for $T > T_\text{c}$, we obtain altogether three
crossings which at higher temperature merge into
a single one.}
\label{self_energy_re}
\end{figure}

{\em Pseudogap regime}.
With the appearance of $\Gamma_{0} \neq 0$  above $T_\text{c}$,
the real part of the self-energy becomes a continuous
function of $\omega$ (see figure~\ref{self_energy_re}).
Consequently, besides the Bogoliubov modes, we now
obtain an additional crossing located in-between. Figure~\ref{poles_num} shows the representative dispersion
curves obtained for $1.1T_\text{c}$, $1.5T_\text{c}$, $2T_\text{c}$ and
compared with the superconducting state (dotted line).
We observe either the three branches or just the single one
at sufficiently high temperatures when the spectral function
$A({\bf k},\omega)$ evolves to a single peak structure.

As some useful example, let us study the Fermi momentum
${\bf k}_\text{F}$, when
(\ref{constraint_for_poles}) simplifies to
\begin{eqnarray}
 \omega \left( 1 -  \frac{ \Delta^{2} }
 { \omega^{2} + \Gamma_{0}^{2}} \right) = 0.
\label{constraint_at_kF}
\end{eqnarray}
In this case, we obtain:
a) two symmetric quasiparticle energies at $\omega_{\pm}
=\pm \tilde{\Delta}$, where $\tilde{\Delta} \equiv
\sqrt{\Delta^{2}-\Gamma_{0}^{2}}$, and b) the in-gap state
at $\omega_{0}=0$. The corresponding imaginary parts
$\Gamma_{\bf k}(\omega)$ are
\begin{eqnarray}
\Gamma_{\bf k}(\omega_{\pm}) & = &
\Gamma_{1}+\Gamma_{0}\,,
\label{imag_pm} \\
\Gamma_{\bf k}(\omega_{0}) & = &
\Gamma_{1}+\frac{\Delta^{2}}{\Gamma_{0}}\, .
\label{imag_0}
\end{eqnarray}
Since $\Gamma_{1}$ does not vary above $T_\text{c}$, the
temperature dependence of $\Gamma^{-1}_{\bf k}(\omega_{i})$
is controlled by $\Gamma_0$. Using the experimental estimations
\cite{Norman-98a,Norman-98b}, we thus find the qualitatively opposite
temperature variations of $\Gamma_{\bf k}(\omega_{\pm})$
and $\Gamma_{\bf k}(\omega_{0})$ shown in figure~\ref{fig2}.
These quantities correspond to the life-times
of quasiparticles and, therefore, we conclude that:
\begin{itemize}
\item[{a)}] in-gap quasiparticles are forbidden for
the superconducting state due to vanishing
$\Gamma^{-1}_{\bf k}(\omega_{0}) = 0$
(in other words, spectrum consists of just the
Bogoliubov modes typical of the BCS theories),
\item[{b)}] in the pseudogap state above $T_\text{c}$,
where $\Delta\neq 0$ and $\Gamma_{0}\neq 0$, besides
the Bogoliubov branches there emerge in-gap states
 which initially at $T_\text{c}^{+}$ represent the heavily
 overdamped modes.
\end{itemize}

\begin{figure}[!t]
\epsfxsize=8cm\centerline{\epsffile{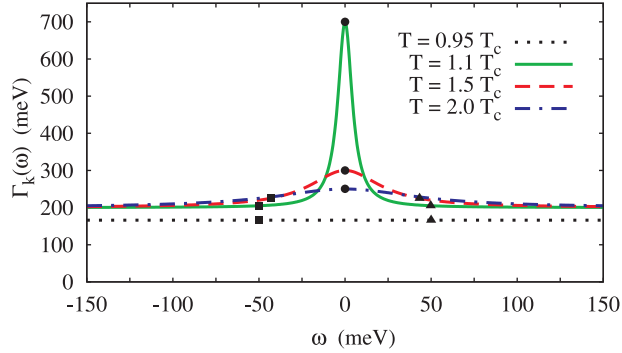}}
\caption{(Color online)
The imaginary part $\Gamma_{\bf k}
(\omega)$ for the same set of parameters as used in figure~\ref{self_energy_re}. The filled symbols indicate the value
of  $\Gamma_{\bf k}(\omega)$ and position of the crossings
$\omega = E_{\bf k}$ of the lower Bogoliubov mode
(squares), in-gap state (circles) and the upper Bogoliubov
branch (triangles).}
\label{self_energy_im}
\end{figure}

\begin{figure}[!b]
\epsfxsize=8cm\centerline{\epsffile{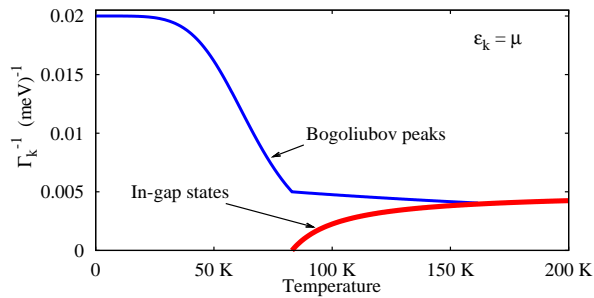}}
\caption{(Color online)
Temperature dependence of
the inverse broadening $\Gamma_{\bf k}^{-1}$ which
corresponds to the effective life-time of the Bogoliubov
modes (thin line) and the in-gap states (thick curve)
obtained for $\varepsilon_{\bf k} = \mu$.}
\label{fig2}
\end{figure}

\begin{figure}[!t]
\epsfxsize=6cm\centerline{\epsffile{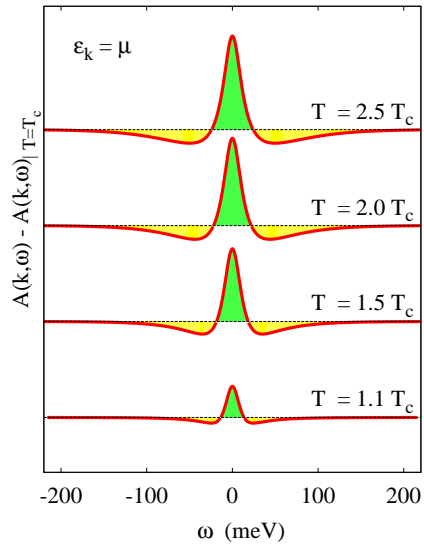}}
\caption{(Color online)
Transfer of the spectral weight from
the Bogoliubov quasiparticle peaks towards the in-gap states
obtained using (\ref{selfenergy}) for $\varepsilon_{\bf k} = \mu$.
Temperature dependence of the total transferred spectral weight
is shown in figure~\ref{weight}.}
\label{transfer}
\end{figure}

\begin{figure}[!b]
\epsfxsize=8cm\centerline{\epsffile{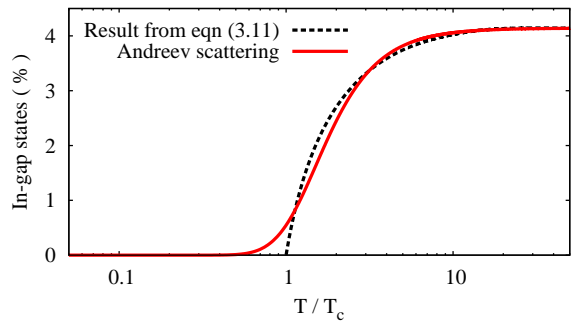}}
\vspace{-2mm}
\caption{(Color online)
Spectral weight
corresponding to the in-gap states obtained from
the phenomenological ansatz (\ref{selfenergy}) for
$\varepsilon_{\bf k} = \mu$ (dotted curve) and
solution of the toy model  (\ref{BFM}) for
$\varepsilon_{0} = 0 = E_{0}$  (solid line).}
\label{weight}
\end{figure}

At a first glance, our conclusions seem to be in conflict
with the ARPES data, which have not reported any
pronounced in-gap features. Nevertheless, various
studies of the pseudogap clearly revealed a rather
negligible temperature dependence of  $\Delta(T)$
upon passing $T_\text{c}$ (at least for the anti-nodal areas).
Instead of closing this gap,  the low energy states
are gradually filled-in \cite{STM}. Such a behavior can
be thought as an indirect signature of the in-gap states,
which for increasing temperatures absorb more and
more spectral weight. To support this conjecture,
we illustrate in figure \ref{transfer} an ongoing
transfer of the spectral weights between the
Bogoliubov quasiparticles and in-gap states. Using
 (\ref{selfenergy}),
we show  the spectral function $A({\bf k},\omega)$
subtracting its value at $T_\text{c}$ in analogy to the
detailed experimental discussion by T. Kondo et al.
\cite{Kondo2009}. In-gap states
emerge around $\omega_{0}$ and gradually gain
the spectral weight (figure~\ref{weight})
simultaneously increasing their life-time.

Intrinsic broadening of the in-gap states \cite{Romano-10}
unfortunately obscures their observation by the spectroscopic
tools at temperatures close to $T_\text{c}$. These states  might be,
however, probed  indirectly. T.~Senthil and P.A.~Lee \cite{Senthil-09a,Senthil-09b}
suggested that such states could be responsible for
the magnetooscillations observed experimentally
by N.~Doiron-Leyraud et al. \cite{Doiron-07a,Doiron-07b}.
They indicated that pair-coherence extending only over short
spatial- and temporal length naturally implies the pair decay
(scattering) into the in-gap fermion states. This line of
reasoning has been also followed by some other groups
\cite{Romano-10,Micklitz-09}.

\section{Microscopic toy model}

Pairing of the cuprate superconductors occurs on a local
scale, practically between the nearest neighbor lattice sites.
To account for an interplay between the paired and unpaired
charge carriers taking place in the pseudogap regime we
consider here the following simplified picture
\begin{eqnarray}
\hat{H}_\text{loc} = \varepsilon_{0} \sum_{\sigma}
\hat{c}_{\sigma}^{\dagger} \hat{c}_{\sigma}
+E_{0} \hat{b}^{\dagger} \hat{b} +
\left( \Delta \hat{b}^{\dagger} \hat{c}_{\downarrow}
\hat{c}_{\uparrow} + \mbox{\rm h.c.} \right),
\label{BFM}
\end{eqnarray}
where $\hat{c}_{\sigma}^{(\dagger)}$ correspond to the unpaired
fermions and $\hat{b}^{(\dagger)}$ to the pairs (hard-core bosons).
We assume that in the pseudogap state, neither the fermions
nor the hard-core boson pairs are long-living because of their
mutual scattering by the Andreev charge exchange term.
The same type of scattering, although in the momentum
space, has been considered  in reference
\cite{Senthil-09a,Senthil-09b} within the lowest order diagrammatic
treatment. On a microscopic footing, the Hamiltonian
(\ref{BFM}) can be regarded as the effective low energy
description of the plaquettized Hubbard model \cite{Auerbach-09a,Auerbach-09b}.

Neglecting the itinerancy of the charge carriers,
we can obtain a rigorous solution for a given local
cluster (not to be confused with the individual copper
sites in CuO$_{2}$ planes \cite{Ranninger-10}). Exact
diagonalization of the Hilbert space yields the following
single particle Green's function \cite{Domanski2011}
\begin{eqnarray}
G(\omega) = \frac{{\cal{Z}}_\text{QP}}{\omega - \varepsilon_{0}}
+\frac{1 - {\cal{Z}}_\text{QP}}{\omega - \varepsilon_{0} -
\frac{|\Delta|^{2}}{\omega+\varepsilon_{0}-E_{0}}} .
\label{local}
\end{eqnarray}
Let us notice that the second term on rhs of
(\ref{local}) acquires the same structure as imposed
by (\ref{selfenergy}). In the present case, no imaginary
terms appear but the structure of the Green's function
(\ref{local}) mimics the role of $\Gamma_{0}$. Formally,
it describes the bonding and antibonding states
originating from the Andreev scattering and
besides that we also have a remnant of the non-interacting
propagator  $\left[ \omega - \varepsilon_{0}\right]^{-1}$
whose spectral weight is given by ${\cal{Z}}_\text{QP}$.

The quasiparticle weight ${\cal{Z}}_\text{QP}$ depends
on occupancies of the fermion and boson levels.
As an example, we  explore
here the symmetric (i.e., half-filled) case with
$\varepsilon_0 = 0$ and $E_{0} = 0$ when
${\cal{Z}}_\text{QP} = {2}/[3+\cosh({{|\Delta|}/{k_\text{B}T}})]$ .
Assuming the typical ratio $|\Delta|/k_\text{B}T_\text{c} = 4$, we plot
in figure~\ref{weight} the temperature dependence of the
unpaired states contribution ${\cal{Z}}_\text{QP}$ to the spectrum.
We find a very good agreement between our simple treatment
and the estimations using the self-energy (\ref{selfenergy}).
It means that the parameter $\Gamma_{0}$ introduced
in references \cite{Norman-98a,Norman-98b} and the local Andreev-type scattering
considered here account for the very same
particle-hole processes inducing the in-gap states.
Transfer of the spectral weight from the paired to
unpaired states (figure~\ref{weight}) confirms
the qualitative agreement over a broad temperature
region and the indication for the same critical point.

For more realistic comparison of the present study  with
the experimental data \cite{Kondo2009}, one obviously has to
consider the itinerant charge carriers.   As a natural
improvement of the local solution (\ref{local}) we
would  expect the following type of Green's function
\begin{eqnarray}
G({\bf k},\omega) = \frac{{\cal{Z}}_\text{QP}({\bf k})}{\omega
 -  \varepsilon_{\bf k}} +  \sum_{\bf q} \frac{
\left[1 - {\cal{Z}}_\text{QP}({\bf k}) \right] f({{\bf q},{\bf k}})}
{\omega - \varepsilon_{\bf k}  - \frac{|\Delta_{\bf k}|^{2}}
{\omega+\varepsilon_{{\bf q}-{\bf k}}-E_{\bf q}}}\,,
\label{extension}
\end{eqnarray}
where $T$-dependent coefficients $f({{\bf q},{\bf k}})$
should be determined via the many-body techniques.
Approaching $T_\text{c}$ from above the predominant
influence comes from ${\bf q} \rightarrow {\bf 0}$
bosons and then we notice that (\ref{extension})
reduces to the ansatz (\ref{selfenergy}). Such results
have been recently reported from the dynamical mean field
calculations for the Hubbard model \cite{Imada2011,Sakai2015}.

\section{Conclusions and outlook}

We have shown that the pairing ansatz (\ref{selfenergy}),
widely used for fitting the experimental ARPES profiles,
above $T_\text{c}$ corresponds to the pair scattering
inducing the single particle fermion states inside the
pseudogap.
Temperature dependent  phenomenological parameter
$\Gamma_{0}$ is found to control the transfer of
the spectral weight from the Bogoliubov quasiparticles
to the unpaired in-gap states.
To model such a process on a microscopic level, we have
considered the scenario in which the local pairs are scattered
into single fermions via the Andreev conversion
\cite{Senthil-09a,Senthil-09b,Ranninger-10,Auerbach-09a,Auerbach-09b}. We have found
a unique relation between the transferred spectral weight
(from the paired to unpaired quasiparticles)
with the non-bonding state $\cal{Z}_\text{QP}$.
It would be instructive to extend the present analysis
onto the case of ${\bf k}$-dependent energy gap. Such a
problem would be closely related to the issue of Fermi arcs,
i.e., partially reconstructed pieces of the Fermi surface,
and to nontrivial angular dependence of the pseudogap
\cite{Kaminski2014}.

\section*{Acknowledgements}
Author is indebted for the fruitful discussions with Adam Kami\'nski,
Roman Micnas, Julius Ranninger, and Karol~I.~Wysoki\'nski.
This work is supported by the National Science Centre in Poland
through the projects DEC-2014/13/B/ST3/04451.

\ukrainianpart

\title{Стани квазічастинок, керованих розсіюванням \\ на попередньо сформованих електронних парах}
\author{T. Доманьский}
\address{Інститут фізики, Університет Марії Кюрі-Складовської,
              20-031 Люблін, Польща}

\makeukrtitle

\begin{abstract}
Проаналізовано еволюцію спектру збудження однієї частинки слабо легованих купратних надпровідників поблизу  антинодальної зони,  беручи до уваги температури вищі і нижчі за фазовий перехід. Досліджено феноменологічну самоенергію, яка відтворює дані  ARPES (кутової фотоемісійної спектроскопії). Показано, що при температурах, вищих  за критичну, така процедура передбачає перехід спектральної ваги від квазічастинок типу Боголюбова до загасаючих станів всередині щілини. Окрім цього подано певні мікроскопічні аргументи, які пояснюють такий процес.
\keywords флуктуаціїї надпровідності, квазічастинки Боголюбова, псевдощілина
\end{abstract}

\end{document}